\documentstyle[12pt]{article}
\input{epsf}
\begin{document}
\rightline{RUB-TPII-9/97}
\rightline{hep-ph/9709436}
\vspace{.3cm}
\begin{center}
\begin{large}
{\bf Nucleon structure functions from the instanton vacuum: \\[.2cm]
Leading and non-leading twists$^\dagger$} \\[1cm]
\end{large}
\vspace{1.4cm}
{\bf M.V.\ Polyakov}$^{\rm a, b}$ {\bf and C. Weiss}$^{\rm b}$  
\\[1.cm]
$^{a}$ {\em Petersburg Nuclear Physics Institute, Gatchina,
St.Petersburg 188350, Russia} 
\\
$^{b}$ {\em Institut f\"ur Theoretische Physik II,
Ruhr--Universit\"at Bochum, \\ D--44780 Bochum, Germany} 
\end{center}
\vspace{1.5cm}
\begin{abstract}
\noindent
We review the description of nucleon structure functions in the instanton
vacuum. This includes the calculation of the twist--2 parton distributions
at a low normalization point as well as higher--twist matrix elements. The
instanton vacuum with its inherent small parameter, the packing fraction of
the instanton medium, $\bar\rho / R$, provides a consistent picture of the
non-perturbative gluon degrees of freedom at the scale 
$\bar\rho^{-1} \simeq 600 \, {\rm MeV}$. The twist--2 quark and antiquark
distribution are of order unity, while the twist--2 gluon distribution is
of order $(\bar\rho / R)^4$. Twist--4 matrix elements determining power
corrections to the Bjorken, Ellis--Jaffe and Gross--Llewellyn-Smith sum
rules are found to be of order $(\bar\rho / R)^0$. We present numerical
estimates for the parametrically large quantities.
\end{abstract}
\vfill
\rule{5cm}{.15mm} \\
{\footnotesize $\dagger$ Talks presented at the XXXVII Cracow School of
Theoretical Physics, Zakopane, May 30 -- June 10, 1997.}
\newpage
In these talks we give a brief summary of recent progress in understanding 
the deep--inelastic structure of the nucleon, both at leading and 
non-leading twist level, in the instanton vacuum
\cite{DPW95,DPPPW96,PP96,BPW97}. Our aim is
to show that the instanton vacuum, with its inherent small parameter ---
the packing fraction of the instanton medium, $\bar\rho / R$ --- provides 
a basis for a consistent and quantitative description of nucleon 
structure functions.
\par
\underline{Leading and non-leading twist.} The non-perturbative information 
which enters in the QCD description
of deep--inelastic scattering and other related experiments is contained 
in nucleon matrix elements of operators of twist 2 and higher.
The moments of the non-power suppressed part of the structure 
functions are given by matrix elements of operators of leading twist; 
in the unpolarized case these are the twist--2 quark and gluon 
operators
\begin{eqnarray}
\lefteqn{i^{n - 1} \sum_{\rm spin} \; \langle P | \; \bar\psi 
\; \tau_{NS, S} \;
\gamma_{\left\{\mu_1 \right.}
\nabla_{\mu_2} \ldots \nabla_{\left. \mu_n \right\}} 
\psi \; \Bigr|_{\mu} \; | P \rangle } && \nonumber \\
&=& 2 \; A^{(n)}_{NS, S} (\mu ) \;
\left[ P_{\mu_1} \ldots P_{\mu_n} - \mbox{traces} \right] ,
\label{O_q_non} \\ 
\lefteqn{ i^{n - 1} \sum_{\rm spin} \; \langle P | \; 
F_{\left\{\mu_1 \alpha \right.} D_{\mu_2} \ldots D_{\mu_{n - 1}} 
F^\alpha_{\left. \;\;\mu_n \right\}} \; \Bigr|_{\mu} \; | P \rangle 
} \nonumber &&
\nonumber \\
&=& 2 \; A^{(n)}_{G} (\mu) \; 
\left[ P_{\mu_1} \ldots P_{\mu_n} - \mbox{traces} \right] ,
\label{O_g_non} 
\end{eqnarray}
and similarly for the polarized case, see \cite{SV82}.
Here, $\tau_S = 1, \tau_{NS} = \tau^3$ are flavor matrices.
Alternatively, one may work with non-local (light-cone) operators, which 
serve as generating functions of the series of local twist--2 operators.
In a partonic language, the matrix elements of these operators can directly 
be interpreted as parton distribution functions\footnote{Here, $q_f(x)$
corresponds to the quark distribution at positive $x$, and to minus the
antiquark distribution at negative $x$.}, the scale dependence of 
which is described by the DGLAP evolution equation:
\begin{eqnarray}
\lefteqn{ q_f (x, \mu) \;\; = \;\;
\int\limits_{-\infty}^\infty \frac{d\lambda}{2\pi} 
e^{i\lambda x} \; \sum_{\rm spin} } && \nonumber \\ 
&& \langle P | \, \bar\psi_f (0) \, {n\hspace{-.5em}/\hspace{.15em}} 
\; \left\{ \mbox{P}\exp
\left[ -i \int_0^\lambda d\lambda' \; n \cdot A (\lambda' n)  
\right] \right\}
\psi_f (\lambda n) \; \Bigr|_{\mu} \, | P \rangle ,
\nonumber \\
\label{q_df}
\\
\lefteqn{ g (x, \mu) \;\; = \;\; 
\int\limits_{-\infty}^\infty \frac{d\lambda}{2\pi} 
e^{i\lambda x} \; \sum_{\rm spin} } && \nonumber \\
&& n_\mu n_\nu
\langle P | \, F_{\mu\alpha} (0) \; \left\{ \mbox{P}\exp
\left[ -i \int_0^\lambda d\lambda' \; n \cdot A (\lambda' n) 
\right] \right\}
F^\alpha_{\;\;\nu}  (\lambda n) \; \Bigr|_{\mu} \, | P \rangle ,
\nonumber \\
\label{g_df}
\end{eqnarray}
where $n$ denotes a light-like four--vector, $n^2 = 0$.
Operators of higher twist arise in the description of power 
corrections \cite{SV82}. For example, in the unpolarized case the
$1/Q^2$--power corrections to the Gross--Llewellyn--Smith sum 
rule are governed by the matrix element of the twist--4, spin--1 
operator ($M_N$ is the nucleon mass)
\begin{eqnarray}
{\textstyle{\frac{1}{2}}} \sum_{\rm spin}
\langle P | \bar\psi \gamma_\alpha \gamma_5 \widetilde{F}^{\beta\alpha} 
\psi | P \rangle &=& 2 M_N^2 \; C^{(2)}_S \; P^\beta .
\label{c2_def}
\end{eqnarray}
In polarized scattering the $1/Q^2$--power 
corrections to the isovector and isosinglet combinations of the first moment 
of the polarized structure function $g_1$ --- the Bjorken and 
Ellis--Jaffe sum rules --- involve the matrix elements of the
twist--3, spin--2 operators \cite{EhrMS94}
\begin{eqnarray}
\lefteqn{ \langle P S | 
\bar\psi \; \tau_{NS, S} \; \left( \gamma^\alpha \widetilde{F}^{\beta\gamma}  
+ \gamma^\beta \widetilde{F}^{\alpha\gamma}  \right) \psi  
| P S \rangle - \mbox{traces}} && 
\label{d2_def} \\
&=& 2 M_N \; d^{(2)}_{NS, S} \; \left[ 2 P^\alpha P^\beta S^\gamma
- P^\gamma P^\beta S^\alpha - P^\alpha P^\gamma S^\beta
+ (\alpha \leftrightarrow \beta ) - \mbox{traces} \right] ,
\nonumber 
\end{eqnarray}
(the same matrix elements contribute also at leading twist level
to the third moment of the structure function $g_2$), and the matrix 
elements of the twist--4, spin--1 operators,
\begin{eqnarray}
\langle P S | \bar\psi \; \tau_{NS, S} \;
\gamma_\alpha \widetilde{F}^{\beta\alpha}
\psi | P S \rangle &=& 2 M_N^3 \; f^{(2)}_{NS, S} \; S^\beta .
\label{f2_def}
\end{eqnarray}
Here, $S$ is the nucleon polarization vector, $S^2 = -1$.
The operators here are assumed to be normalized at scale $\mu$; the 
scale dependence of the matrix elements is described by the renormalization
group equation. The twist--2 quark--, antiquark-- 
and gluon distributions at a low normalization point have been determined
by fits to data from a variety of 
experiments \cite{MRS95,GRV95}. The twist--3 matrix element
$d^{(2)}$ has recently been extracted from measurements of the structure 
function $g_2$ \cite{Abe_p}.
Experimental knowledge of the higher--twist matrix elements 
entering only in power corrections, such as the twist--4 matrix element 
$f^{(2)}$, is still rather poor \cite{JM97}.
\par
Any attempt to calculate the matrix elements mentioned here
from first principles requires an understanding of the non-perturbative 
effects giving rise to the structure of the nucleon. While the gross 
features of the leading--twist quark distributions can be understood in 
phenomenological models like the quark model or the bag model,
to describe the gluon distribution or higher--twist matrix 
elements explicitly involving the gluon field one needs a theory 
of the non-perturbative fluctuations of the gluon field. 
Lattice calculations of structure functions have been making steady 
progress during the last years; however, they are still far from giving 
a satisfactory quantitative description \cite{Goeckeler96}. 
\par
\underline{Instanton vacuum.}
A microscopic picture of the non-perturbative fluctuations of the gluon
field is provided by the instanton vacuum. For an introduction to 
the instanton vacuum and its applications to hadronic physics we refer 
to the extensive literature on this subject, {\em e.g.} the recent 
reviews \cite{SchSh96,D96_Varenna}; we can here touch upon only those
aspects directly relevant to structure functions. Instantons and
antiinstantons ($I$ and $\bar I$ for short) are particular field 
configurations which are solutions of the Euclidean Yang--Mills 
equations, characterized by a size, $\rho$, center, $z$ and color 
orientation given by an $SU(N_c)$ matrix, ${\cal U}$, 
\begin{eqnarray}
A_\mu^a (x; \; z , {\cal U})_{I (\bar I)} &=& f_\nu (x - z) \; O^{ab} 
(\eta^\mp )^b_{\mu\nu} , 
\nonumber \\
f_\nu (x) &=& \frac{2 \rho^2}{(x^2 + \rho^2) x^2} x_\nu, 
\hspace{1cm} O^{ab} \;\; = \;\; 
{\textstyle{\frac{1}{2}}} {\rm tr}\, [\lambda^a {\cal U} 
\lambda^b {\cal U}^\dagger ] .
\label{A_inst}
\end{eqnarray}
Here, $(\eta^\mp )^b_{\mu\nu} = \bar\eta^b_{\mu\nu}, \eta^b_{\mu\nu}$
are the 't Hooft symbols.
Instantons have many special properties; not all of them are of interest
to us here. Let us note only that a 
single $I (\bar I)$ is an $O(4)$--symmetric field configuration --- this
fact makes for important differences between instanton contributions
to operators of different spin, see below.
\par
In the instanton vacuum one considers non-perturbative effects
due to field configurations with a finite density of $I$'s and $\bar I$'s. 
The medium of $I$'s and $\bar I$'s stabilizes itself due to instanton 
interactions \cite{DP84}, meaning that the average size of the 
instantons in medium is finite,
\begin{eqnarray}
\bar\rho &\simeq& (600 \, {\rm MeV})^{-1} .
\end{eqnarray}
The coupling constant is fixed at a scale of order $\bar\rho^{-1}$, 
so when we evaluate matrix elements of QCD operators below it is
implied that the operators are normalized at $\mu \simeq \rho^{-1}$.
It should be stressed that no external scale is introduced here; 
all parameters of the instanton medium are obtained in terms of the 
QCD scale parameter, $\Lambda_{\rm QCD}$. Hence this
approach preserves the essential renormalization properties of QCD.
\par
The most important property of the instanton 
vacuum is the small packing fraction of the medium, {\em i.e.}, the small 
ratio of the average size of the instantons in the medium to the average 
separation between nearest neighbors, ${\bar\rho}/{\bar R} \simeq 1/3$
\cite{Sh82,DP84}. This small parameter is the starting point for a 
systematic analysis of non-perturbative phenomena in this picture.
\par
In particular, the instanton vacuum explains the dynamical breaking of 
chiral symmetry. The Dirac operator in the
background of one $I (\bar I)$ has a localized zero mode,
\begin{eqnarray}
\left[ i {\partial\hspace{-.5em}/\hspace{.15em}} 
+ {A\hspace{-.65em}/\hspace{.3em}}_{I (\bar I)} 
(x; z , {\cal U} ) \right]
\Phi_\pm (x; z , {\cal U} )  = 0.
\end{eqnarray}
In the medium the zero modes associated with the individual
instantons delocalize \cite{DP86}, resulting in a finite fermion
spectral density at zero eigenvalue, which by the Banks--Casher theorem
is equivalent to the chiral condensate \cite{BC80}. Alternatively, one 
can derive the effective action of fermions in the instanton medium in 
the $1/N_c$--expansion, integrating over the instanton coordinates in the 
ensemble \cite{DP86_prep}. An individual $I (\bar I)$ interacts with the 
fermion field through the zero mode, {\em i.e.}, through a ``potential''
\begin{eqnarray}
V_{I(\bar I)} [\bar\psi_f , \psi_f ] &=&
\int d^4 x \int d^4 y \;
\bar\psi_f (x) \; {\partial\hspace{-.5em}/\hspace{.15em}} 
\Phi_\pm (x; z , {\cal U} ) \bar\Phi_\pm (y; z , {\cal U} ) \;
{\partial\hspace{-.5em}/\hspace{.15em}} \psi_f (y) .
\nonumber \\
\label{V_I}
\end{eqnarray}
In leading order in $\bar\rho / R$ the 
effective action exhibits a many--fermionic interaction, which is given 
by the one--instanton average of eq.(\ref{V_I}) and has the form of 
the 't Hooft determinant in flavor indices \cite{tH76},
\begin{eqnarray}
\int d^4 z \, d{\cal U} \;
\prod_f^{N_f} V_{I(\bar I)} [\bar\psi_f , \psi_f ] 
&\propto& \det_{fg} \; \bar\psi_f \, \frac{1 \pm \gamma_5}{2} \, 
\psi_g .
\label{tHooft}
\end{eqnarray}
In addition, there is a form factor (not written here for brevity)
related to the finite size of the zero--mode wave function, which makes the 
interaction vanish for quark momenta larger than $\bar\rho^{-1}$.
Chiral symmetry is spontaneously broken due to this many--fermionic 
interaction: the quarks acquire a dynamical (momentum--dependent) mass, 
a quark condensate develops, and a massless pion appears as a 
collective excitation. In the $1/N_c$--expansion one can easily 
construct the effective action in the chirally broken phase.
It can be formulated as a theory of massive quarks, interacting with
the pion field in a chirally invariant way,
\begin{eqnarray}
Z &=& \int {\cal D} \pi \int {\cal D}\bar\psi {\cal D}\psi\,
\; \exp \int d^4 x \; 
i \bar\psi (x) \; \left[ i {\partial\hspace{-.5em}/\hspace{.15em}} + i M 
e^{i \gamma_5 \tau^a \pi^a (x)}
\right] \; \psi (x) . 
\nonumber \\
\label{effective_theory}
\end{eqnarray}
The effective theory applies for quark momenta up to the 
inverse instanton size, $\bar\rho^{-1} \simeq 600 \, {\rm MeV}$, which 
acts as a cutoff. For the applicability of this effective
theory it is crucial that the ratio of the dynamically generated quark 
mass to the cutoff is proportional to the packing fraction of the 
instanton medium, 
\begin{eqnarray}
M\bar\rho &\propto& \left( \frac{\bar\rho}{R} \right)^2 .
\end{eqnarray}
Hence the diluteness of the instanton medium guarantees that the picture 
of massive ``constituent'' quarks applies in a parametrically wide range
of momenta. Finally, we note that the nucleon is obtained in the 
$1/N_c$--expansion as a chiral soliton of the effective theory, 
eq.(\ref{effective_theory}) \cite{DPP88}. This picture of the nucleon
gives a very reasonable description of a variety of hadronic
properties such as the $N$--$\Delta$ splitting,
electromagnetic formfactors, axial coupling constants {\em
etc.}\ \cite{Review}.
\par
The instanton vacuum, with the resulting effective chiral theory,
allows to evaluate hadronic matrix elements of QCD operators
involving the gluon field. In ref.\cite{DPW95} a method was developed
by which ``gluonic'' operators can systematically be represented as
effective operators in the effective chiral theory.
Let ${\cal F}[A]$ be a gluonic operator, {\em i.e.}, some function
of the gauge field. To leading order in the packing fraction, 
$\bar\rho / R$, the interaction
of the gluon operator with the fermion field is mediated by single 
instantons. The effective operator, denoted by 
$\mbox{``${\cal F}$''}$, is 
obtained by substituting in ${\cal F}[A]$ the gauge field of one 
$I (\bar I)$ and integrating over the collective coordinates,
\begin{eqnarray}
\mbox{``${\cal F}$''}[\bar\psi , \psi ] &\propto& \sum_{I + {\bar I}}
\int d^4 z \, d{\cal U} \; {\cal F} \left[ \, A_{I (\bar I)} (z, U) \, 
\right] \; \prod_f^{N_f} V_{I(\bar I)} [\bar\psi_f , \psi_f ] .
\label{effop}
\end{eqnarray}
In higher orders of $\bar\rho /R$ one needs to take 
into account many-instanton contributions to the effective operator. 
Below we shall need the effective operator corresponding
to the gauge field itself, which is given by 
({\em cf.}\ eq.(\ref{A_inst})) \cite{BPW97}
\begin{eqnarray}
\mbox{``$A$''} (x)_\mu^a &\propto& 
\sum_{\pm} \; \int d^4 z \; f_\nu (x - z) \;
\bar\psi (z) \; \frac{\lambda^a}{2} \sigma_{\mu\nu} 
\frac{1\pm \gamma_5}{2} \; e^{\pm i \pi^a (z) \tau^a} \; \psi (z) .
\nonumber \\
\label{A_eff}
\end{eqnarray}
Again we have suppressed the form factors coming from the zero mode
wave function. Note the presence of the pion field, as a result of which 
the effective operator eq.(\ref{A_eff}) is chirally invariant --- as it 
should be, since the gluon field is flavor neutral. 
\par
It is important to note that
the representation of QCD operators as effective operators
is possible relying entirely on the approximations already inherent 
in the effective theory --- the diluteness of the instanton medium
and the $1/N_c$--expansion; no additional assumptions are required.
It was shown in \cite{DPW95} that this approach preserves the essential
renormalization properties of QCD; for example, the QCD trace and $U(1)$ 
anomalies are realized at the level of hadronic matrix elements. 
This method is thus well suited for computing matrix 
elements of the QCD operators of twist 2 and higher twist 
of interest here.
\par
\underline{Twist--2 matrix elements in the instanton vacuum.}
Let us first consider the twist--2 gluon operators, 
eq.(\ref{O_g_non}). Since we are interested only in forward 
matrix elements we may average the operator position over the 4--volume.
The definition of the effective operator, eq.(\ref{effop}), 
implies the integral over the instanton coordinate, $z$. Since the 
instanton field is $O(4)$ symmetric the only tensor one has at hands to 
construct the effective operator is the Kronecker delta, but from it is 
impossible to construct a traceless symmetric tensor! (When working
with light--like components this follows from $\delta_{++} = 0$.) Thus, 
the effective operators
for the twist--2 gluon operators, eq.(\ref{O_g_non}), or, more
generally, the nonlocal operator, eq.(\ref{g_df}), 
vanish at one--instanton level. We conclude that 
the twist--2 gluon distribution is parametrically suppressed in
$\bar\rho / R$. 
\par
To determine the twist--2 gluon distribution quantitatively one has 
to include (at least) the two--instanton contribution to the effective 
operator. Preliminary results indicate that the gluon distribution
is, in fact, proportional to $(M\bar\rho )^2$, which is parametrically 
of order $(\bar\rho / R)^4$.\footnote{The gluon distribution in the
instanton vacuum has recently been studied in a different approach 
by Kochelev \cite{Kochelev97}. However, the contributions taken into 
account there do not represent the full answer to order 
$(\bar\rho / R)^4$.} It is interesting to note that, numerically, 
a suppression of the gluon relative to the singlet quark momentum fraction
by a factor $(M\bar\rho )^2 \simeq 0.3$ is consistent with the GRV
parametrization of the data at a normalization point of 
$\mu \simeq 600\,{\rm MeV}$ \cite{GRV95}.
\par
In the twist--2 quark operators, eq.(\ref{O_q_non}), 
the gauge field enters through the covariant derivative, or, equivalently,
through the path--ordered exponential in the non-local operator, 
eq.(\ref{q_df}). It is interesting to ask how much the gauge field 
contributes 
to the moments of the quark distribution functions in the instanton vacuum, 
which is formulated in the so-called singular gauge in which the instanton
field has the form eq.(\ref{A_inst}). For simplicity, let us consider the 
second moment of the singlet unpolarized quark distribution, $A^{(2)}_S$, 
which is given by the matrix element of the operator
\begin{eqnarray}
i \langle P | \bar\psi \gamma_{\left\{\mu_1 \right.}
\left( \partial_{\left. \mu_2 \right\}}  - i \frac{\lambda^a}{2} 
A_{\left. \mu_2 \right\}}^a \right) \psi | P \rangle - \mbox{traces} 
&=& 2 A^{(2)}_S P_{\mu_1} P_{\mu_2} - \mbox{traces} .
\nonumber \\
\label{M1}
\end{eqnarray}
It is instructive to compute the matrix element not immediately in the 
nucleon, but first in a ``constituent'' quark, {\em i.e.}, the massive 
quark of the effective chiral theory. It is easy to see that the short 
derivative in 
eq.(\ref{M1}) makes an order unity contribution to $A^{(2)}_S$. Computing 
the gauge field contribution to the matrix element using the effective 
operator, eq.(\ref{A_eff}), one finds that it is of order 
$(M\bar\rho )^2 \propto (\bar\rho / R)^4$, {\em i.e.}, parametrically
suppressed relative to the short derivative. Thus we have
\begin{eqnarray}
A^{(2)}_{S, {\rm quark}} &=& 1 \; + \; O\left[ \left(\frac{\rho}{R} 
\right)^4 \right] .
\end{eqnarray}
This is consistent with the fact that the gluon distribution is of 
higher order in the packing fraction: To order $(\bar\rho / R)^0$ the 
quarks carry the entire momentum, and the gluon distribution is zero.
(To compute the full $(\rho / R)^4$ contribution would, again, require to
take into account two--instanton contributions to the effective operators,
as well as to the effective quark action.)
The above statements are easily generalized to higher moments. In fact, 
speaking of parton distribution functions ``inside the constituent 
quark'' one may say that
\begin{eqnarray}
\sum_f^{N_f} q_f(x)_{\rm quark} &=& \delta (x - 1) \; + \; 
O\left[ \left( \frac{\bar\rho}{R} \right)^4 \right] , \\
g(x)_{\rm quark} &=& 
O\left[ \left( \frac{\bar\rho}{R} \right)^4 \right] .
\end{eqnarray}
To leading order in $\bar\rho / R$ the constituent quark has no structure.
Consequently, in leading order in $\bar\rho / R$ it is justified to
identify the distribution of ``constituent'' quarks and antiquarks in the 
nucleon with the actual parton distribution at the scale 
$\mu \simeq \bar\rho^{-1}$. In another way of saying, when computing the 
twist--2 quark distribution function in the effective theory 
in leading order in $\bar\rho / R$ one can identify the QCD quark fields 
normalized at $\mu \simeq \bar\rho^{-1}$ with the quark fields of the 
effective chiral theory and put the path--ordered exponential in 
eq.(\ref{q_df}) to unity. This is the ``quarks--antiquarks only''
approximation which was employed to compute the quark and antiquark 
distributions of the nucleon in the chiral quark soliton 
model \cite{DPPPW96}. 
We have thus seen that this approximation has a parametric justification 
in the instanton vacuum.
%
%
\begin{figure}[t]
\epsfxsize=15cm
\epsfysize=12cm
\epsffile{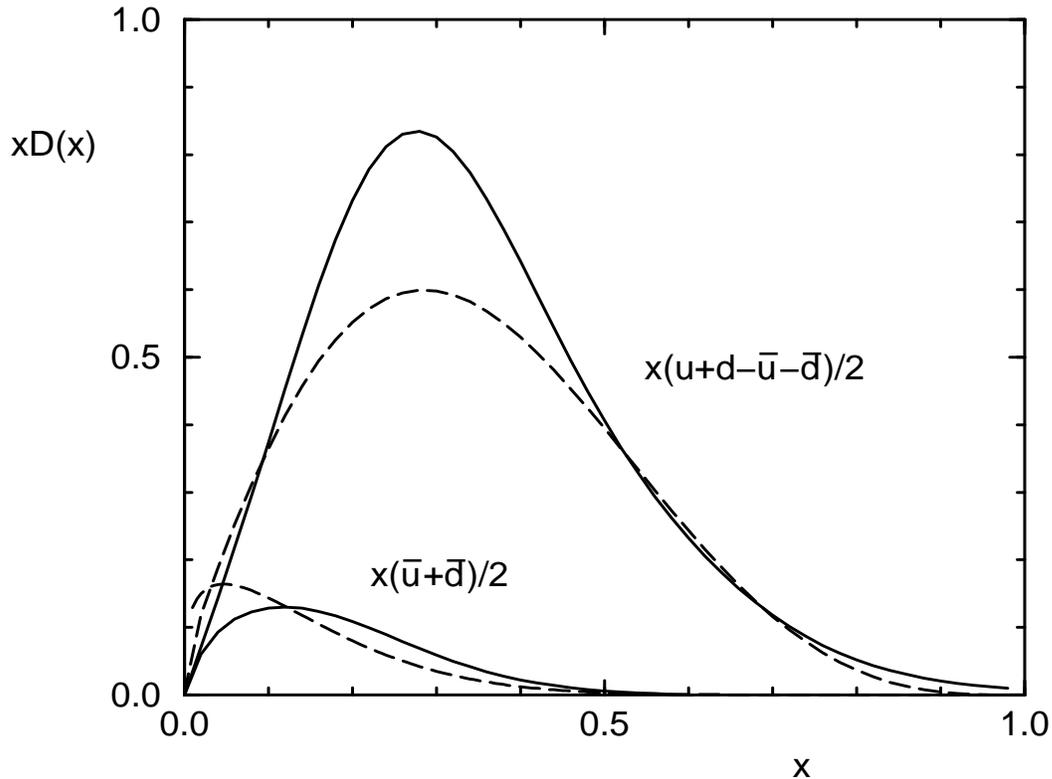}
\caption[]{{\em Solid lines}: The isosinglet unpolarized valence quark and 
antiquark distributions computed in the chiral soliton model of 
the nucleon \cite{DPPPW96}. {\em Dashed lines}: NLO--parametrization 
of GRV ($\mu^2 = 0.34 \, {\rm GeV}^2$) \cite{GRV95}.}
\label{fig_1}
\end{figure}
\par
The twist--2 quark and antiquark distributions of the nucleon computed 
in the chiral quark 
soliton model satisfy all general requirements, such as positivity, 
proper normalization {\em etc}. This is a fully field--theoretic description
of the nucleon, which, in particular, makes possible a consistent calculation 
of the antiquark distributions. All partonic sum rules (baryon number, 
isospin, momentum, Bjorken sum rule) are satisfied within the model.
Computed so far were the leading distribution functions in the large--$N_c$
limit, the isosinglet unpolarized and isovector polarized 
\cite{DPPPW96}, as well as the isovector transverse polarized 
distribution \cite{PP96}. The isosinglet unpolarized quark and antiquark
distributions are shown in Fig.\ref{fig_1}, together with the 
GRV parametrization \cite{GRV95}; for the polarized distributions see 
the original papers.
\par
To order $(\bar\rho / R)^0$ the constituent quarks have no
structure, and the constituent quarks and antiquarks carry the 
entire nucleon momentum. As a result the singlet quark distribution
calculated in this approximation is generally larger than the GRV 
parametrization, which includes gluons at the low normalization point, 
see Fig.1. In higher orders of $\bar\rho / R$ one starts to systematically
resolve the structure of the constituent quark in terms of the original 
QCD degrees of freedom, and the nucleon momentum gets distributed among 
quarks and gluons at the low scale. A 30\% gluon 
momentum fraction at the low scale obtained by GRV \cite{GRV95} is 
consistent with the gluon distribution in the instanton vacuum being 
suppressed by a factor $(M\bar\rho)^2 \simeq 0.3$.
\par
To summarize the discussion of twist--2 operators, one may say that at
twist--2 level the effects of the instanton medium are essentially 
contained in the dynamical quark mass generated in the dynamical 
breaking of chiral symmetry. Instanton contributions to the twist--2
operators, in the sense of effective operators, eq.(\ref{effop}),
are parametrically suppressed. This is what one could call a 
``constituent'' quark picture. We note that, contrary to other approaches 
where the ``constituent'' quark is a largely philosophical object, 
in the instanton vacuum this term has a well--defined meaning, thanks to 
the parameter $\bar\rho /R$ \cite{D96}.
\par
\underline{Higher--twist matrix elements in the instanton vacuum.}
From the above discussion one may have the impression that in the instanton
vacuum ``gluonic'' contributions to operators are always parametrically 
suppressed relative to quark operators. This is not so --- instantons can 
make order $(\bar\rho / R)^0$ in operators of {\em higher twist}. The most
immediate way to convince oneself of this is to consider a particular 
higher--twist operator whose matrix elements vanishes by the QCD equations 
of motion. In addition, this exercise provides a beautiful check for the 
consistency of the effective 
operator method. Consider the twist--4 matrix element
obtained by projecting the operator in eq.(\ref{M1}) not on spin two 
but on spin zero:
\begin{eqnarray}
i \langle P | \bar\psi {\nabla\hspace{-.65em}/\hspace{.3em}} 
\psi | P \rangle 
&\equiv&
\langle P | \bar\psi \left( i {\partial\hspace{-.5em}/\hspace{.15em}} + 
{A\hspace{-.65em}/\hspace{.3em}} \right) 
\psi | P \rangle \;\; = \;\; 0.
\label{eom_qcd}
\end{eqnarray}
In QCD the matrix element is zero by virtue of the QCD equations of motion. 
In the instanton vacuum we find, computing the contribution of the gauge 
field in eq.(\ref{eom_qcd}) using the effective operator, 
eq.(\ref{A_eff}), one may show that
\begin{eqnarray}
\langle P | \bar\psi \left( i {\partial\hspace{-.5em}/\hspace{.15em}} + 
\mbox{``${A\hspace{-.65em}/\hspace{.3em}}$''}
\right) \psi | P \rangle 
&=& 
\langle P | \bar\psi \left( i {\partial\hspace{-.5em}/\hspace{.15em}} 
+ i M e^{i \gamma_5 \tau^a \pi^a (x)} \right) 
\psi | P \rangle \;\; = \;\; 0.
\label{eom_inst}
\end{eqnarray}
(The calculation is actually rather involved; see \cite{BPW97} for details).
Here the gauge field contribution to the operator is of order unity, 
{\em i.e.}, of the same order as that of the short derivative. As a
result the QCD operator of eq.(\ref{eom_qcd}) reduces to an
effective operator which vanishes identically due to the equations
of motion of the effective chiral theory. From this we learn
two things: First, instantons can make order unity contributions in twist--4
operators. Second, the method of effective operators
preserves a principal feature of QCD: matrix elements of operators which
are zero in QCD due to the QCD equations of motion are automatically zero
in the effective theory. We note that also other operators whose forward 
matrix elements vanish in QCD,
\begin{eqnarray}
\langle P | \bar\psi \gamma_\alpha F^{\beta\alpha}
\psi | P \rangle , \hspace{2em}
\langle P S | \bar\psi \gamma_\alpha \gamma_5 F^{\beta\alpha}
\psi | P S \rangle ,
\end{eqnarray}
give zero matrix elements when translated to the effective chiral theory.
%
%
\noindent
\begin{table}[t]
\begin{center}
\[
\begin{array}{|l|c|c|c|c|c|c|c|}
\hline
& f^{(2)}_{NS} & d^{(2)}_{NS} & C^{(2)}_{S} & \mbox{scale}/{\rm GeV}^2 \\
\hline
\mbox{Instanton vacuum} \; \cite{BPW97}
& -0.10  & \sim 10^{-3} & 0.36 & \sim 0.4 \\
\mbox{Sum rules} \; \cite{BBK90}  
&  -0.20  &       0.072  & \mbox{---}  & 1   \\
\mbox{Sum rules} \; \cite{Stein95t4}      
& -0.072 &        0.072  & \mbox{---}  & 1   \\
\mbox{Bag model} \; \cite{JiU94}         
&  0.11  &        0.063  & \mbox{---}   & 5 \\
\mbox{Lattice} \; \cite{Goeckeler96} 
& \mbox{---}  &  -0.13   & \mbox{---}   & 4 \\
\mbox{Sum rules} \; \cite{BK87}
& \mbox{---}  & \mbox{---} & 0.37 & 1 \\
\mbox{E142, E143, E154} \; \cite{Abe_p}                    
& \mbox{---}  &  0.043 \pm 0.046 & \mbox{---} & 3 \\
\mbox{Power corr.} \; \cite{JM97}         
& 0.10 \pm 0.28  & \mbox{---} & \mbox{---}  & 1 \\
\hline
\end{array}
\]
\end{center}
\caption[]{Numerical results for the flavor--nonsinglet
spin--dependent twist--4 and 3 matrix elements $f^{(2)}_{NS}$ and 
$d^{(2)}_{NS}$, eqs.(\ref{d2_def}, \ref{f2_def}), and the 
flavor--singlet spin--independent twist--4 matrix element $C^{(2)}_{S}$, 
eq.(\ref{c2_def}). Shown are the results obtained from the instanton 
vacuum \cite{BPW97}, from QCD sum rule calculations 
\cite{BBK90,Stein95t4,BK87}, from the bag model \cite{JiU94}, 
and from lattice calculations \cite{Goeckeler96}.
Also shown are estimates of $d^{(2)}_{NS}$ based on 
measurements of the structure function $g_2$ \cite{Abe_p}, and 
estimates of $f^{(2)}_{NS}$ from an analysis of power corrections
to $g_1$ \cite{JM97}.}
\label{table_numeric}
\end{table}
\par
We can now turn to the calculation of the matrix elements of twist--3
and 4 operators appearing in power corrections, 
eqs.(\ref{c2_def}, \ref{d2_def}, \ref{f2_def}). Again, the qualitative
features can be seen by studying the matrix elements in ``constituent''
quark states. Computing the matrix elements of the effective operators
corresponding to eqs.(\ref{c2_def}, \ref{d2_def}, \ref{f2_def}) one finds 
that twist--3 and 4 matrix elements are of different order in the packing 
fraction:
\[
\begin{array}{lrclcl}
\mbox{twist 4:} & f^{(2)}, C^{(2)} 
&\sim& (M\bar\rho )^0 &\sim& 
{\displaystyle \left(\frac{\bar\rho}{R}\right)^0 } , \\[.3cm]
\mbox{twist 3:} & d^{(2)} &\sim& (M\bar\rho )^2 \log M\bar\rho 
&\sim& {\displaystyle
\left(\frac{\bar\rho}{R}\right)^4 \log \left(\frac{\bar\rho}{R}\right) .}
\\
\end{array}
\]
Again, the reason for this can be seen in the $O(4)$ symmetry of the
individual $I (\bar I)$. Note that the parametric order of the matrix elements 
is determined by two factors: {\em i)} the number of instantons participating 
in the effective operator (here we have included only the one--instanton
contribution), and {\em ii)} the dependence of the quark loop integrals
in the matrix element (obtained by closing the quark lines on the
many--fermionic effective operator) on the cutoff, $\bar\rho^{-1}$, 
keeping in mind that $(M\bar\rho )^2 \sim (\bar\rho /R )^4$. An order 
unity contribution can come only from integrals which are ``quadratically 
divergent'', meaning they are proportional to $\bar\rho^{-2}$. 
Incidentally, this last fact implies that, in the framework of our effective 
theory, the dominant contributions to higher--twist matrix elements come 
from ``divergent'' loop diagrams where the effective many--fermionic operator 
couples to a {\em single} constituent quark, not from diagrams 
describing interactions of more than one constituent quark mediated 
by the many-fermionic effective operator. In the latter all momenta are 
cut by the bound--state wave function of the nucleon, not by the
cutoff, $\bar\rho^{-1}$. Losely speaking, one may thus say
that in our picture higher--twist matrix elements measure 
properties of the individual constituent quarks, not correlations between 
them.
\par
We have computed the leading higher--twist nucleon matrix elements in 
the large--$N_c$ limit; see \cite{BPW97} for details. These are the flavor 
nonsinglet 
spin--dependent ones ($d^{(2)}_{NS}, f^{(2)}_{NS}$) and flavor singlet 
spin--independent ones ($C^{(2)}_{S}$). Results are shown in 
table \ref{table_numeric}. The result for $d^{(2)}_{NS}$ should be taken 
as an order--of--magnitude estimate; to compute it accurately at 
level $(\bar\rho / R)^4$ one needs to include the two--instanton 
contribution to the effective 
operator. Note that the results for $f^{(2)}$ and $C^{(2)}$ agree well 
with estimates from QCD sum rules \cite{BBK90,Stein95t4,BK87}; our value 
for $d^{(2)}$ is consistent with estimates based on measurements of the 
structure function $g_2$ \cite{Abe_p}.
\par
\underline{Summary.} The predictions of the instanton vacuum 
for nucleon matrix elements relevant to structure functions can 
be summarized as follows:
\begin{itemize}
\item 
At twist--2 level, the quark and antiquark distributions
are of order unity in the packing fraction. In this order they can
be computed in the effective chiral theory without including instanton 
contributions to the twist--2 operators, {\em i.e.}, replacing covariant 
by short derivatives in eq.(\ref{O_q_non}), or dropping the path--ordered 
exponential in the light-cone operator, eq.(\ref{q_df}).
The twist--2 gluon distribution is of order $(\bar\rho / R)^4$; to compute 
it one needs to take into account at least two-instanton contributions to 
the effective operators.
\item 
The instanton vacuum implies a {\em hierarchy of twists}: 
Large ---  that is, $(\bar\rho / R)^0$ --- contributions
are found in operators of lowest spin (twist--4, in our case), while 
the contributions to operators of higher spin (twist--3 and 2) are 
suppressed. The reason for this pattern is the $O(4)$--symmetry of the 
single instanton.
\end{itemize}
The instanton vacuum provides a consistent framework for describing
the non-perturbative input necessary for a complete understanding
of DIS experiments. The key element, which makes possible
a systematic approach to non-perturbative phenomena, is the 
small parameter $\bar\rho / R$ inherent in this picture.
The ``quarks--antiquarks only'' approximation for twist--2 operators,
in connection with the chiral quark soliton model of the nucleon, 
gives a very successful description of the twist--2 quark and antiquark 
distributions of the nucleon, both polarized and 
unpolarized \cite{DPPPW96}. As to higher twists, one may hope
that increasing accuracy of the measurements of polarized and 
unpolarized structure functions (power corrections) or, possibly,
semi-inclusive measurements, will allow to test the specific predictions 
of the instanton vacuum more accurately.
\par
Much remains to be done on the theoretical side. In particular, one should 
refine the effective operator approach
to be able to compute also parametrically small matrix elements, first of 
all the twist--2 gluon distribution. In addition to taking into account
the two--instanton contributions to the effective operator this 
requires to compute also the effective quark action to higher orders
in $M\bar\rho$, since many properties --- for example, the correct 
realization of the QCD equations of motion --- depend on the 
consistency of the definitions of effective operators and the effective
action. Work in this direction is in progress.
\par
It is a pleasure to thank J. Balla, D.I.\ Diakonov, V.Yu.\ Petrov, and 
P.V.\ Pobylitsa for a stimulating collaboration and many interesting 
discussions, as well as K. Goeke for encouragement and multiple support.
\par
The work reported here has been supported in part by the Deutsche 
Forschungsgemeinschaft (DFG), by a joint grant of the DFG and the Russian 
Foundation for Basic Research, and by COSY (J\"ulich). 
M.V.P.\ is supported by the A.v.Humboldt Foundation. 
\newpage
\end{document}